\documentclass[a4paper,10pt,twoside]{cpc-hepnp}

\usepackage{multicol}
\usepackage{graphicx}
\usepackage{booktabs}
\usepackage{amssymb,bm,mathrsfs,bbm,amscd}
\usepackage[tbtags]{amsmath}
\usepackage{lastpage}

\begin{document}

\fancyhead[co]{\footnotesize Author1~ et al: Instruction for typesetting manuscripts}

\footnotetext[0]{}

\title{A study of meson-meson potential in the chiral quark model}

\author{
\quad LI Ming-Tao$^{1,2;1)}$\email{limt@mail.ihep.ac.cn}%
\quad DONG Yu-Bing$^{1,2;2)}$\email{dongyb@mail.ihep.ac.cn}%
\quad ZHANG Zong-Ye$^{1,2;3)}$\email{zhangzy@mail.ihep.ac.cn} }
\maketitle

\address{%
$^1$ Institute of High Energy Physics, Chinese Academy of Sciences, Beijing 100049, China\\
$^2$ Theoretical Physics Center for Science Facilities (TPCSF), CAS,
Beijing 100049, China\\}

\begin{abstract}
An effective potential in a meson-meson system is discussed based on
the SU(3) chiral constituent quark model, and the analytic form of
the potential is explicitly given. In addition, the effective
potential is employed to study the bound state problem of
$\omega\phi$, which is related to the new resonance of $f_0(1810)$
observed in BESII very recently.

\end{abstract}



\begin{multicols}{2}

\section{Introduction}
Recently, many new hadronic states have been discovered, such as
light new resonances of X(1835) and $f_0(1810)$ observed by BESII,
and the new charmonium or charmonium-like states of X(3872),
Y(3940), $Z^+(4430)$ et al. measured in Belle and Babar
\cite{bepc1,bepc2,new1,new2}. Since it is difficult to accommodate
those new resonances in the conventional quark model, and many of
them locate just below the threshold of two mesons, some
interpretations, like the molecular picture and tetraquark states
\cite{mol1,mol2}, which are different from the conventional
quark-antiquark constituent quark model, have been proposed to
understand their structures.

Since most of the interpretations in the literature are based on the
effective Lagrangians in the hadronic level, a more sophisticated
understanding of the new resonances in the quark model is required.
We know that the SU(3) chiral constituent quark model is one of the
most successful quark models which can well reproduce the
nucleon-nucleon, nucleon-hyperon interactions and baryon
spectroscopy simultaneously ~\cite{zhangzy1,zhangzy3}. The model
considers the one-boson exchange including the scalar and
pseudoscalar mesons, and its Lagrangian is constrained by the chiral
symmetry. In the most calculations based on the SU(3)
chiral constituent quark model for the baryon-baryon interactions
~\cite{zhangzy1,zhangzy3} and even for the new resonances
~\cite{zhanghx, wang2}, the RGM or GCM methods are often employed,
where the total system is regarded to be a composite system with two
clusters and the obtained effective potential is expressed in terms
of generator-coordinates. Here we derive a new effective potential
between the two clusters in another way with the chiral quark model.
Namely, we try to express the effective potential directly in terms
of the relative coordinate between the two clusters other than the
generator-coordinates. It is expected that the newly obtained
potential is a more realistic one. Then, we apply this effective
potential to study the bound state problem of a $\omega\phi$ system,
which is closely related to the new resonance of $f_0(1810)$
recently observed in the BESII experiments ~\cite {bepc2}.

This paper is organized as follows. In section 2, the analytical
effective potential for a meson-meson system is derived based on the
SU(3) chiral constituent quark model with one-boson exchange. The
numerical result of the effective potential for the $\omega\phi$
system is given in section 3. A discussion of the bound state
problem of the $\omega\phi$ system with this potential is given in
the last section.
\section{Analytical effective potential for a meson-meson
system}

For a system with two mesons, in cluster model, we only need
consider the interaction between different clusters, that is, two
different mesons. As a consequence, the sum of that kind of
contributions between different clusters gives our total effective
potential (see Fig.1 for an illustration).
\begin{center}
\includegraphics[width=4cm]{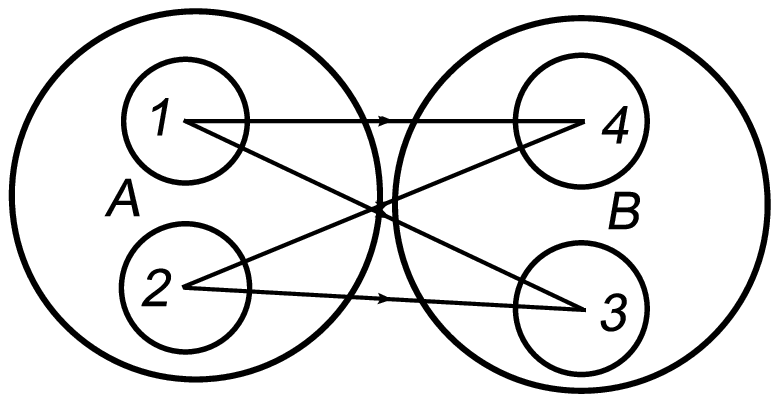}
\figcaption{\label{fig1}   Interactions between clusters }
\end{center}

The total Hamiltonian of such a meson-meson system is
\begin{eqnarray}
H=T+V.
\end{eqnarray}
Here, we have neglected the interaction potentials between two
quarks within each cluster. $T$ is the kinetic energy operator:
\begin{eqnarray}
T=\sum_iT_i-T_{cm}
\end{eqnarray}
and $V$ is the potential operator which indicates the interactions
between the two mesons:
\begin{eqnarray}
V&=&\sum_{i\in A,j\in B}V_{ij}, \\
V_{ij}&=&V^{OGE}(r_{ij})+V^{conf}(r_{ij})+V^{ch}(r_{ij}),
\end{eqnarray}
while $V^{OGE}(r_{ij}),V^{conf}(r_{ij})$ and $V^{ch}(r_{ij})$ are
respectively one-gluon-exchange potential, confinement potential and
one-meson-exchange potential
between $i$-th quark in cluster A and $j$-th quark in cluster
B£¬with $i=(1,2)$ and $j=(3,4)$. The forms of one-gluon-exchange
potential and confinement potential have been shown in Ref.
\cite{wang1}.

The chiral Lagrangian of the quark-quark interaction under the SU(3)
chiral quark model is
\begin{eqnarray}
\mathcal{L}^{ch}_I=-g_{ch}\emph{F}(\emph{\textbf{q}}^2)\overline{\psi}
(\sum^8_{a=0}\lambda_a\sigma_a+i\gamma_5\sum^8_{a=0}\lambda_a\pi_a)\psi.
\end{eqnarray}
Here, $\lambda^a$ indicates the Gellman flavor matrix, $\sigma^a$ indicates
the scalar masons, $\pi^a$ stands for pseudoscalar masons, and
$\emph{F}(\emph{\textbf{q}}^2)$ expresses the form factor of the chiral
field with the form of
\begin{eqnarray}
\emph{F}(\emph{\textbf{q}}^2)=(\frac{\Lambda^2}{\Lambda^2+\emph{\textbf{q}}^2})^{1/2}.
\end{eqnarray}
In non-relativistic limit, we get the quark-quark interaction in
momentum space. After the Fourier transformation, we reach the
potential in coordinate space. From Ref. \cite{zhangzy3}, we know
the central part of the potential with the  scalar meson exchanges
is
\begin{eqnarray}
V^{\sigma_a}_{cen}(r_{ij})&=&-\frac{g^2_{ch}}{4\pi}\frac
{\Lambda^2{m_a}}{\Lambda^2-{m^{2}_a}}
Y_1(r_{ij})\lambda^a_i\lambda^a_j, \label{eq:s0}
\end{eqnarray}
and the one with the pseudoscalar meson exchange is
\begin{eqnarray}
V^{\pi_a}_{cen}(r_{ij})&=&{\frac{g^2_{ch}}{4\pi}}{\frac
{\Lambda^2m^{3}_a}{\Lambda^2-{m^{2}_a}}}
\dfrac{1}{m_im_j}\nonumber \\
&&\times
\frac{1}{12}Y_3(r_{ij})(\sigma_i\cdot\sigma_j)\lambda^a_i\lambda^a_j,
\label{eq:s1}
\end{eqnarray}
where $m_i$ and $m_j$ denote the masses of $i$-th quark in cluster A
and $j$-th quark in cluster B. $\lambda^a(a=0,...,8)$ in eqs.
(\ref{eq:s0},\ref{eq:s1}) indicate flavor matrices correspondence to
scalar mesons $\sigma^a$ or pseudoscalar mesons $\pi^a$. In the two
equations we have
\begin{eqnarray}
Y_1(r_{ij})&=&Y(m_a r_{ij})-{\frac{\Lambda}{m_a}}Y
(\Lambda r_{ij}),\label{eq:y1}\\
Y_3(r_{ij})&=&Y(m_a r_{ij})-({\frac{\Lambda}{m_a}})^3 Y(\Lambda
r_{ij}),\label{eq:y2}
\end{eqnarray}
and the Yukawa function is
\begin{eqnarray}
Y(x)=\frac{1}{x}e^{-x}.
\end{eqnarray}
Here, $m_a$ denotes the meson mass in scalar or pseudoscalar nonets.
Now the chiral potential
\begin{eqnarray}
V^{ch}(r_{ij})=\sum_a V^{\sigma_a}(r_{ij})+\sum_a
V^{\pi_a}(r_{ij}).\label{eq:-1}
\end{eqnarray}

The effective potentials in eqs.(\ref{eq:s0},\ref{eq:s1}) have three
parts: orbital, associated with $r$; spin, associated with
$\sigma_i\cdot\sigma_j$ which only appears in the pseudoscalar meson
case in eq.(\ref{eq:s1}); flavor, associated with $\lambda^a_i
\lambda^a_j$. Except the orbital part, the other two parts are easy
to deal with. As a result, we focus our attention on the orbital one
in this paper.

Defining the Jacobi coordinates as follows
\begin{eqnarray}
\left\{\begin{array} {r@{\quad=\quad}l}
\overrightarrow{\xi_1} & \overrightarrow{r_1}-\overrightarrow{r_2}\\
\overrightarrow{R_1} & \frac{m_1\overrightarrow{r_1}+m_2\overrightarrow{r_2}}{m_1+m_2}\\
\overrightarrow{\xi_2} & \overrightarrow{r_3}-\overrightarrow{r_4}\\
\overrightarrow{R_2} &
\frac{m_3\overrightarrow{r_3}+m_4\overrightarrow{r_4}}{m_3+m_4},
\end{array}\right.
\label{eq:10}
\end{eqnarray}
where $\overrightarrow{\xi_1}$ and $\overrightarrow{\xi_2}$ are
respectively the relative coordinates within the two clusters, and
$\overrightarrow{R_1}$ and $\overrightarrow{R_2}$ the center-of-mass
coordinates, the single particle coordinates can then be re-written
in terms of the new Jacobi corrdinates as
\begin{eqnarray}
\left\{ \begin{array}
{r@{\quad=\quad}l}
\overrightarrow{r_1} & \overrightarrow{R_1}+\frac{m_2}{m_1+m_2}\overrightarrow{\xi_1}\\
\overrightarrow{r_2} & \overrightarrow{R_1}-\frac{m_1}{m_1+m_2}\overrightarrow{\xi_1}\\
\overrightarrow{r_3} & \overrightarrow{R_2}+\frac{m_4}{m_3+m_4}\overrightarrow{\xi_2}\\
\overrightarrow{r_4} &
\overrightarrow{R_2}-\frac{m_3}{m_3+m_4}\overrightarrow{\xi_2}.
\end{array}\right.
\label{eq:03}
\end{eqnarray}
Moreover $\overrightarrow{r_{13}}$ can be written as
\begin{eqnarray}
\overrightarrow{r_{13}}=\overrightarrow{\xi}+\frac{m_2}{m_1+m_2}\overrightarrow{\xi_1}-\frac{m_4}{m_3+m_4}\overrightarrow{\xi_2}
\end{eqnarray}
with $\overrightarrow{\xi}=\overrightarrow{R_1}-\overrightarrow{R_2}$ being
the relative coordinates between the
two clusters.

Usually, we take a Guassian-like wavefunction for the orbital
wave-function of each quark in the cluster for simplity. It is
\begin{eqnarray}
\varphi(\textbf{r})&=&(\frac{m_q\omega}{\pi})^{3/4}e^{-\frac{m_q\omega}{2}\textbf{r}^2},
\end{eqnarray}
where \textbf{r} and $m_q$ stands for the coordinate and the mass of
each quark respectively, and the parameter $\omega$ is chosen as
$0.5GeV\thickapprox2.522fm^{-1}$ traditionally. The normalized
orbital wavefunction of cluster A is
\begin{eqnarray}
\varphi_{A_o}=(\frac{m_1\omega}{\pi})^{3/4}(\frac{m_2
\omega}{\pi})^{3/4}e^{-\frac{\omega}{2} (m_1r_1^2+m_2r_2^2)}.
\end{eqnarray}
It is also true for cluster B. Then the total orbital wavefunction
of cluster A and cluster B is
\end{multicols}
 \ruleup
\begin{eqnarray}
|\varphi_{A_o}\varphi_{B_o}\rangle&=&(\frac{m_1\omega}{\pi})^{3/4}(\frac{m_2\omega}{\pi})^{3/4}
(\frac{m_3\omega}{\pi})^{3/4}(\frac{m_4\omega}{\pi})^{3/4}
e^{-\frac{\omega}{2}(m_1r_1^2+m_2r_2^2+m_3r_3^2+m_4r_4^2)}.
\label{eq:04}
\end{eqnarray}
\ruledown \vspace{0.5cm}
\begin{multicols}{2}
The wavefunction (\ref{eq:04}) of the system can be written in terms
of the newly defined Jacobi corrdinates in eq.(\ref{eq:10}) as
\end{multicols}
 \ruleup
\begin{eqnarray}
|\varphi_{A_o}\varphi_{B_o}\rangle&=&(\frac{\mu_{12}\omega}{\pi})^{3/4}(\frac{\mu_{34}\omega}
{\pi})^{3/4}(\frac{M
\omega}{\pi})^{3/4}(\frac{\mu_{12,34}\omega}{\pi})^{3/4}
e^{-\frac{\omega}{2}(\mu_{12}\xi_1^2+\mu_{12}\xi_2^2+MR_c^2+\mu_{12,34}\xi^2)},
\end{eqnarray}
\ruledown \vspace{0.5cm}
\begin{multicols}{2}
where
\begin{eqnarray}
\mu_{12}&=&\frac{m_1m_2}{m_1+m_2},\\ \nonumber
\mu_{34}&=&\frac{m_3m_4}{m_3+m_4},\\ \nonumber
M&=&m_1+m_2+m_3+m_4,\\ \nonumber
\mu_{12,34}&=&\frac{(m_1+m_2)(m_3+m_4)}{m_1+m_2+m_3+m_4},\\
\nonumber
\overrightarrow{\xi}&=&\overrightarrow{R_1}-\overrightarrow{R_2},\\
\nonumber
\overrightarrow{R_c}&=&\frac{m_1\overrightarrow{r_1}+m_2\overrightarrow{r_2}+m_3\overrightarrow{r_3}+m_4\overrightarrow{r_4}}{m_1+m_2+m_3+m_4}.
\end{eqnarray}

The total effective potential should only have  relevance to the
relative coordinates between the two clusters, that is:
\begin{eqnarray}
V(\xi)&=&\sum_{ij}V_{ij}(\xi),\\
V_{ij}(\overrightarrow{\xi})&=&V_{ij}^{OGE}(\xi)+V_{ij}^{conf}(\xi)+V_{ij}^{ch}(\xi),\label{eq:00}
\end{eqnarray}
where $V_{ij}(\xi)$ is the effective interaction between the i-th
quark in cluster A and j-th quark in the cluster B. It is
\begin{eqnarray}
V_{ij}(\xi)=\frac{\langle\varphi_{Ao}\varphi_{Bo}|V(r_{ij})\mathcal{A}|\varphi_{Ao}\varphi_{Bo}\rangle}
{\langle\varphi_{Ao}\varphi_{Bo}|\mathcal{A}|\varphi_{Ao}\varphi_{Bo}\rangle},
\label{eq:01}
\end{eqnarray}
In eq.(\ref{eq:01}), $\mathcal{A}$ is the anti-symmetrization
operator:
\begin{eqnarray}
\mathcal{A}&=&(1-P_{13})(1-P_{24}).
\end{eqnarray}
There are four terms in eq.(\ref{eq:01}), and here we take direct
term as an example for our manipulation£¬ which has no relation with
exchange operator $P_{ij}$.


We see in eqs.(\ref{eq:s0},\ref{eq:s1}) that all the meson exchange
potentials are the algebraic sum of two Yukawa potentials with
different parameters and coefficients, so we just calculate the effective potential with
simple Yukawa form
\begin{eqnarray}
\mathcal{V}(r_{ij})=\frac{e^{-mr_{ij}}}{r_{ij}},
\end{eqnarray}
and then
\begin{eqnarray}
\mathcal{V}_{ij}(\xi)=\frac{\langle\varphi_{Ao}\varphi_{Bo}|\frac{e^{-mr_{ij}}}{r_{ij}}\mathcal{A}|\varphi_{Ao}\varphi_{Bo}\rangle}
{\langle\varphi_{Ao}\varphi_{Bo}|\mathcal{A}|\varphi_{Ao}\varphi_{Bo}\rangle}.
\end{eqnarray}

In the case of $i,j=1,3$ we have
\end{multicols}
\ruleup
\begin{eqnarray}
\mathcal{V}_{13}(\xi)&=&\int d\overrightarrow{\xi_1}\int
d\overrightarrow{\xi_2}\frac{e^{-mr_{13}}}{r_{13}}(\frac{\mu_{12}\omega}{\pi})^{3/2}
(\frac{\mu_{34}\omega}{\pi})^{3/2}e^{-\omega(\mu_{12}\xi_1^2+\mu_{12}\xi_2^2)}\nonumber \\
&=&\int d\overrightarrow{\xi_1}\int
d\overrightarrow{\xi_2}\frac{e^{-m|\xi+\frac{m_2}{m_1+m_2}
\overrightarrow{\xi_1}-\frac{m_4}{m_3+m_4}\overrightarrow{\xi_2}|}}{|\xi+\frac{m_2}{m_1+m_2}
\overrightarrow{\xi_1}-\frac{m_4}{m_3+m_4}\overrightarrow{\xi_2}|}
(\frac{\mu_{12}\omega}{\pi})^{3/2}(\frac{\mu_{34}\omega}{\pi})^{3/2}e^{-\omega(\mu_{12}\xi_1^2+\mu_{12}\xi_2^2)}.
\label{eq:02}
\end{eqnarray}
\ruledown \vspace{0.5cm}
\begin{multicols}{2}
After some manipulations(see Appendix),  and taking
\begin{eqnarray}
\label{eq:05}
\beta&=&\frac{\mu_{12}\mu_{34}\omega}{\mu_{12}(\frac{m_4}{m_3+m_4})^2+\mu_{34}(\frac{m_2}{m_1+m_2})^2},
\end{eqnarray}
we finally reach
\end{multicols}
\ruleup
\begin{eqnarray}
\mathcal{V}_{13}(\xi)&=&\frac{1}{2\xi}e^{\frac{m^2}{4\beta}}\{e^{-m\xi}\{1-erf[-\sqrt{\beta}(\xi-\frac{m}{2\beta})]\}
-e^{m\xi}\{1-erf[\sqrt{\beta}(\xi+\frac{m}{2\beta})]\}\}.\label{eq:08}
\end{eqnarray}
\ruledown \vspace{0.5cm}
\begin{multicols}{2}
Similarly, one can easily deduce the other three $\mathcal{V}_{ij}$,
while the only difference is the value and the form of $\beta$ in
eq.(\ref{eq:05}).

Applying this result to eqs.(\ref{eq:s0}-\ref{eq:y2}), we could get
the orbital part of the meson-exchange potential in direct term in
eq.(\ref{eq:01}). Applying this method, we could get the
one-gluon-exchange potential and confinement potential between the
two clusters. The sum of meson-exchange potential,
one-gluon-exchange potential and confinement potential gives the
contribution of direct term in eq.(\ref{eq:01}). The process to
handle the exchange terms in eq.(\ref{eq:01}) which are associated
with $P_{ij}$ resembles what has been done to the direct term. Other
parts of this potential, like the spin and flavor ones could be
easily obtained (see Ref. ~\cite{wang1}).
\section{The numerical result of the effective
potential for the $\omega\phi$ system}

Here, we will employ our formalism to study the bound state problem
of the $\omega\phi$ system. The flavor part of the wavefunction of
$\omega\phi$ system is
\begin{eqnarray}
|\varphi_{\omega_f}\varphi_{\phi_f}\rangle=|
\frac{1}{\sqrt{2}}(u\overline{u}+d\overline{d})(s,\overline{s})\rangle.
\end{eqnarray}
Because there isn't same flavor quarks in the two different color
singlet clusters,  the one-gluon-exchange potential and confinement
potential don't exist in $\omega\phi$ system. Then, the realistic
effective potential is the total contribution of meson-exchange
potential:
\begin{eqnarray}
V(\xi)=\sum_{ij}\frac{\langle\varphi_{\omega}\varphi_{\phi}|V^{ch}(r_{ij})\mathcal{A}|\varphi_{\omega}\varphi_{\phi}\rangle}
{\langle\varphi_{\omega}\varphi_{\phi}|\mathcal{A}|\varphi_{\omega}\varphi_{\phi}\rangle}.
\label{eq:06}
\end{eqnarray}
Considering the flavor part of $V^{ch}(r_{ij})$ in
eqs.(\ref{eq:s0},\ref{eq:s1},\ref{eq:-1}), we  know that only
certain kinds of meson-exchange can exist in eq.(\ref{eq:06}).
From PDG \cite{pdg}, we know: $m_{K}$=494MeV, $m_{\eta}$=548MeV,
$m_{\eta'}$=958MeV, $m_{\omega}$=782MeV and $m_{\phi}$=1020MeV. To
get a reliable result, we use the following parameters with which
the spectroscopy of baryons can be well fitted
\cite{zhangzy1,zhangzy2}: the scalar meson masses as
$m_{\sigma}$=595MeV, $m_{\epsilon}$=980MeV, $m_{\kappa}$=980MeV, the
cutoff mass as $\Lambda$=1100MeV, the quark masses as $m_u$=313MeV
and $m_s$=470MeV, $\omega$=2.522$fm^{-1}$£¬ and the quark and chiral
field coupling constant $g_{ch}$= 2.621.  After elaborate
investigation, we know that in direct term of eq.(\ref{eq:06}),
there are four pairs of $\sigma$, $\epsilon$, $\eta$ and $\eta^{'}$
exchange respectively as shown in Figs.\ref{fg:f2} with $\sigma$
shown in the short dot line, $\epsilon$ the dash line, $\eta$ the
dot line and $\eta^{'}$ the dash dot line. In exchange terms of
eq.(\ref{eq:06}), there are two pairs of $\kappa$ and $K$ exchange
respectively as shown in Figs.\ref{fg:f3} with $\kappa$ shown in the
short dash line, $K$ the dash dot dot line.
\begin{center}
\includegraphics[width=8cm]{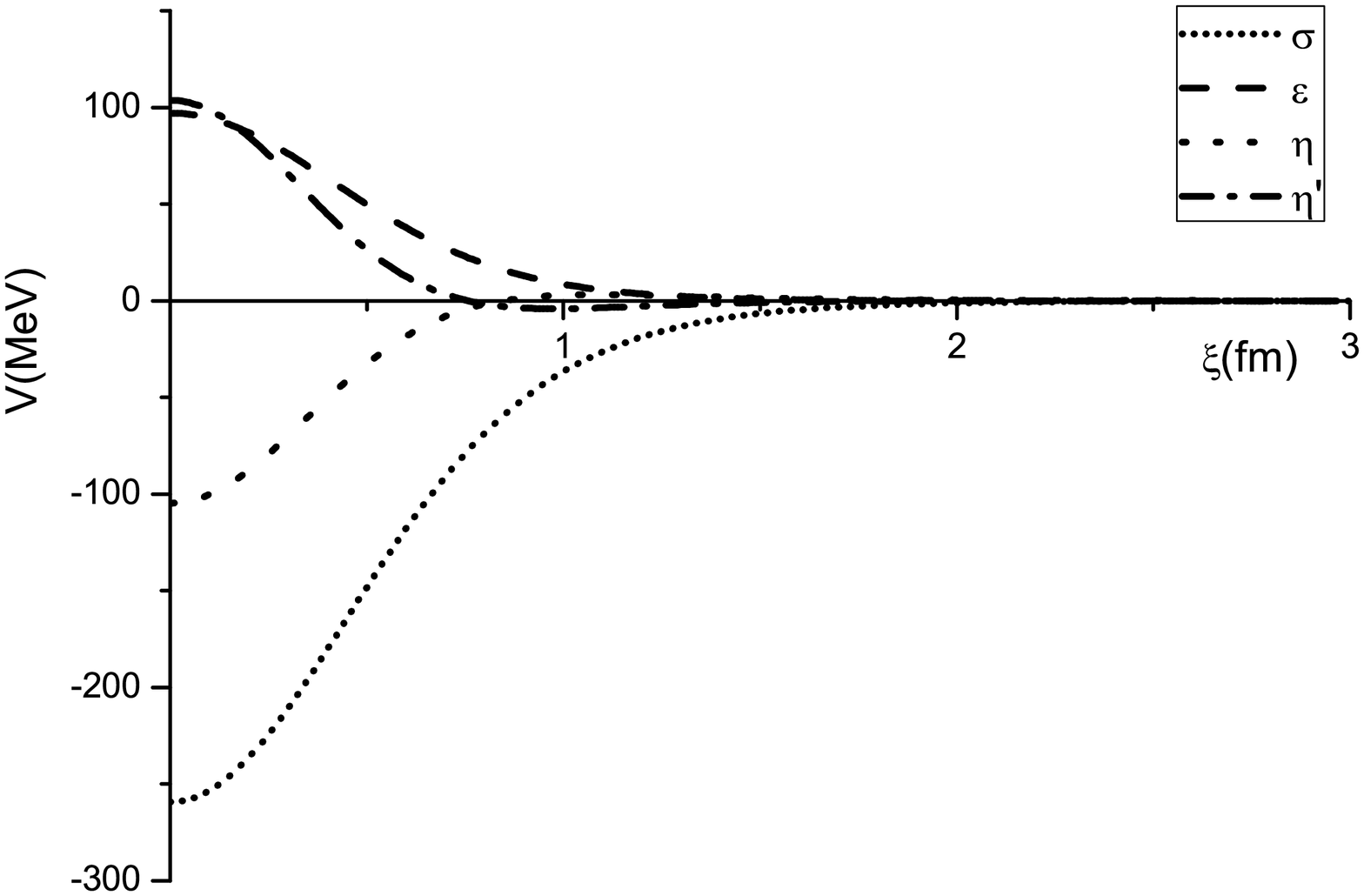}
\figcaption{\label{fig2}   Direct term's contribution to the total
potential } \label{fg:f2}
\end{center}
\begin{center}
\includegraphics[width=8cm]{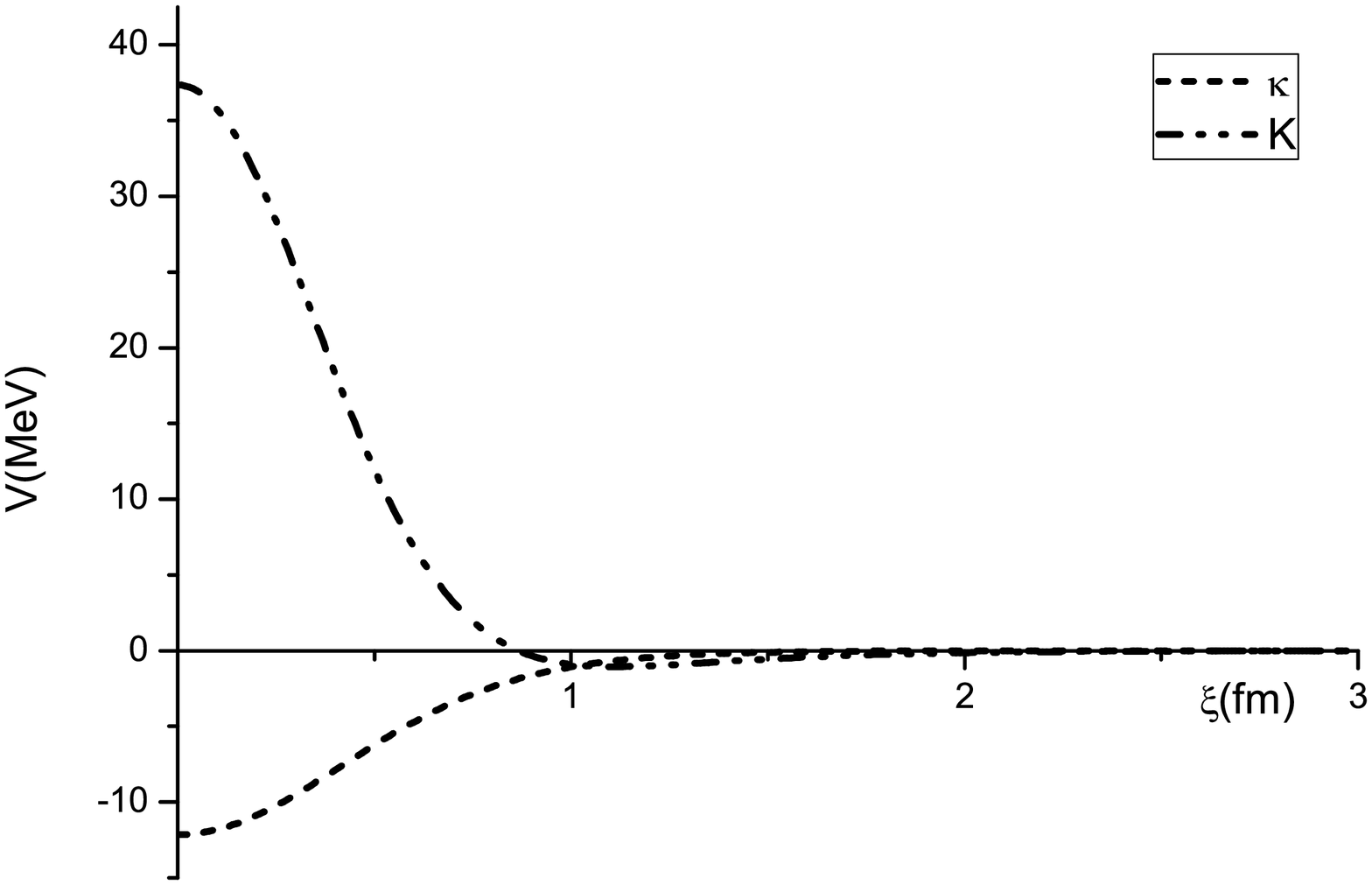}
\figcaption{\label{fig3}   Exchange term's contribution to the total
potential }\label{fg:f3}
\end{center}

From Figs.\ref{fg:f2}-\ref{fg:f3}, we see the contributions of
$\sigma$ exchange, $\eta$ exchange and $\kappa$ exchange are
attractive, while the contributions of $\epsilon$ exchange, $\eta'$
exchange and $K$ exchange are repulsive. These features of the
effective chiral meson exchange potentials are consistent with the
results of RGM calculation~\cite{wang2}.

Now, we can easily determine the total potential as the sum of the
contribution as shown in Figs.\ref{fg:f2}-\ref{fg:f3}. It is shown
in Fig.\ref{fg:f4} with the solid line in comparison with the
different part contributions:

\begin{center}
\includegraphics[width=8cm]{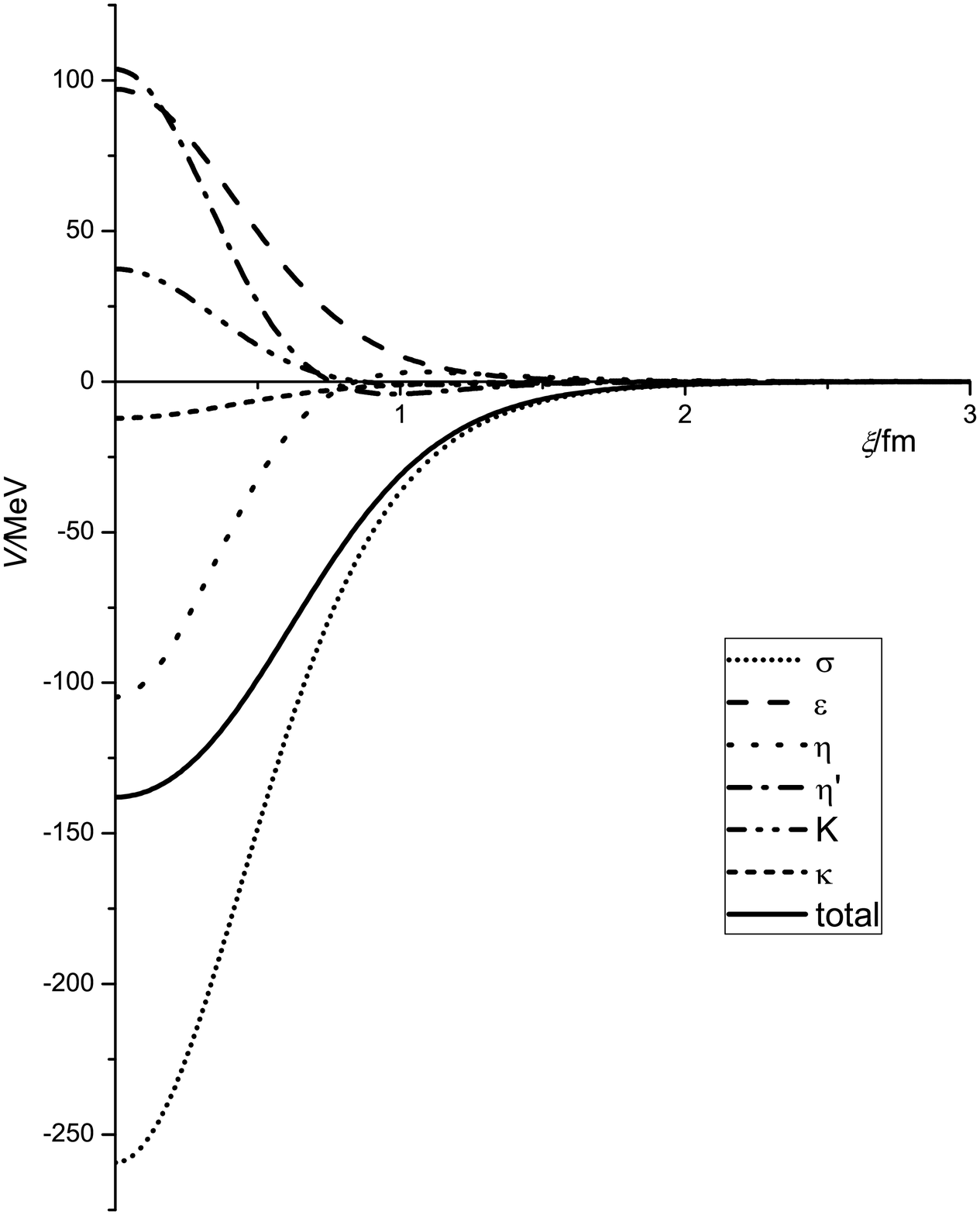}
\figcaption{\label{fg:f4}   The total potential and all the
different parts' contribution}
\end{center}

Finally, we will trace out whether an $\omega\phi$ quasi-bound state
can exist. Since we have deduced the realistic effective potential
which is expressed in terms of relative coordinate between $\omega$
and $\phi$ and chosen the mass of $\sigma$ as 595MeV, we can solve
the Schr\"{o}dinger Equation, and work out the existance of
eigenfunction and eigenvalue. For this purpose we use the computer
program developed by Lucha and Sch\"{o}berl ~\cite{lucha1}.
Unfortunately, we find that there isn't any eigenfunction or
eigenvalue. Then we can draw a conclusion that $\omega$ and $\phi$
can't form a steady system in our approach, which confirms to the
conclusion of Ref. \cite{wang2} We also notice that if the mass of
$\sigma$ is readjusted to 520MeV, this system would have one
eigenstate with an eigenvalue of 0.26MeV.

\section{Summary}

The main purpose of this paper is to deduce an analytical form of
the interaction potential between different clusters in SU(3) chiral
quark model. Compared with the generator-coordinate method in
solving this problem, our method is more realistic.
More importantly, the short-range behavior of the two clusters could
be clearly worked out. This method could be generalized to a more
practical form of the potential which would include the tensor
interaction terms and spin-orbital coupling terms. It could be also
applied to other hadronic molecules and pentaquark systems.

\emph{This work is supported  by the National Sciences Foundations
Nos. 10775146, 10775148, 10975146, and by Ministry of Science and
Technology of China(2009CB825200). The authors thank Professor Franz
F.Sch\"{o}berl for many useful discussions. }

\end{multicols}
\vspace{-1mm} \centerline{\rule{80mm}{0.1pt}} \vspace{2mm}
\begin{multicols}{2}
\subsection*{Appendix}
For eq.(\ref{eq:02}), to further simplify our derivation, we make
another coordinates transformations:
\begin{eqnarray}
\left\{ \begin{array} {r@{\quad=\quad}l}
\overrightarrow{X} & \overrightarrow{\xi}+\frac{m_2}{m_1+m_2}\overrightarrow{\xi_1}-\frac{m_4}{m_3+m_4}\overrightarrow{\xi_2}\\
\overrightarrow{Y} &
\frac{m_3+m_4}{2m_4}\overrightarrow{\xi_1}+\frac{m_1+m_2}{2m_2}\overrightarrow{\xi_2}.
\end{array}\right.
\end{eqnarray}
The inverse transformations are
\begin{eqnarray}
\left\{ \begin{array} {r@{\quad=\quad}l}
\overrightarrow{\xi_1} & \frac{m_4}{m_3+m_4}\overrightarrow{Y}+\frac{m_1+m_2}{2m_2}(\overrightarrow{X}-\overrightarrow{\xi})\\
\overrightarrow{\xi_2} &
\frac{m_2}{m_1+m_2}\overrightarrow{Y}-\frac{m_3+m_4}{2m_4}(\overrightarrow{X}-\overrightarrow{\xi}),
\end{array}\right.
\end{eqnarray}
Then, $\mathcal{V}_{13}$ is
\end{multicols}
\ruleup
\begin{eqnarray}
\mathcal{V}_{13}(\xi)
&=&(\frac{\mu_{12}\omega}{\pi})^{3/2}(\frac{\mu_{34}\omega}{\pi})^{3/2}\int
d\overrightarrow{X}\int
d\overrightarrow{Y}\frac{e^{-mX}}{X}e^{-\mu_{12}\omega[\frac{m_4}{m_3+m_4}
\overrightarrow{Y}+\frac{m_1+m_2}{2m_2}(\overrightarrow{X}-\overrightarrow{\xi})]^2}\nonumber
\\&&
\times e^{-\mu_{34}\omega[\frac{m_2}{m_1+m_2}\overrightarrow{Y}-\frac{m_3+m_4}{2m_4}(\overrightarrow{X}-\overrightarrow{\xi})]^2}\nonumber \\
&=&(\frac{\mu_{12}\omega}{\pi})^{3/2}(\frac{\mu_{34}\omega}{\pi})^{3/2}\int
d\overrightarrow{X}\int d\overrightarrow{Y}\frac{e^{-mX}}{X}
e^{-\mu_{12}\omega[(\frac{m_4}{m_3+m_4})^2Y^2+(\frac{m_1+m_2}{2m_2})^2(\overrightarrow{X}-\overrightarrow{\xi})^2]}\nonumber
\\&&
\times e^{-\mu_{34}\omega[\frac{m_4(m_1+m_2)}{m_2(m_3+m_4)}
\overrightarrow{Y}\cdot(\overrightarrow{X}-\overrightarrow{\xi})]}
e^{-\mu_{12}\omega[(\frac{m_2}{m_1+m_2})^2Y^2+(\frac{m_3+m_4}{2m_4})^2(\overrightarrow{X}-\overrightarrow{\xi})^2]}
e^{\mu_{34}\omega[\frac{m_2(m_3+m_4)}{m_4(m_1+m_2)}
\overrightarrow{Y}\cdot(\overrightarrow{X}-\overrightarrow{\xi})]}\nonumber \\
&=&(\frac{\mu_{12}\omega
}{\pi})^{3/2}(\frac{\mu_{34}\omega}{\pi})^{3/2}\int
d\overrightarrow{X}\int d\overrightarrow{Y}\frac{e^{-mX}}{X}
e^{-\mu_{12}\omega[(\frac{m_4}{m_3+m_4})^2Y^2+(\frac{m_1+m_2}{2m_2})^2(\overrightarrow{X}-\overrightarrow{\xi})^2]}\nonumber
\\&&
\times e^{-\mu_{12}\omega[\frac{m_4(m_1+m_2)}{m_2(m_3+m_4)}
\overrightarrow{Y}\cdot(\overrightarrow{X}-\overrightarrow{\xi})]}
e^{-\mu_{34}\omega[(\frac{m_2}{m_1+m_2})^2Y^2+(\frac{m_3+m_4}{2m_4})^2(\overrightarrow{X}-\overrightarrow{\xi})^2]}
e^{\mu_{34}\omega[\frac{m_2(m_3+m_4)}{m_4(m_1+m_2)}\overrightarrow{Y}\cdot(\overrightarrow{X}-\overrightarrow{\xi})]}\nonumber \\
&=&(\frac{\mu_{12}\omega}{\pi})^{3/2}(\frac{\mu_{34}\omega}{\pi})^{3/2}\int
d\overrightarrow{X}\int d\overrightarrow{Y}\frac{e^{-mX}}{X}
e^{-[\mu_{12}\omega(\frac{m_4}{m_3+m_4})^2+\mu_{34}\omega(\frac{m_2}{m_1+m_2})^2]Y^2}\nonumber \\
&&\times
e^{-[\mu_{12}\omega(\frac{m_1+m_2}{2m_2})^2+\mu_{34}\omega(\frac{m_3+m_4}{2m_4})^2](\overrightarrow{X}-\overrightarrow{\xi})^2}
e^{[-\mu_{12}\omega\frac{m_4(m_1+m_2)}{m_2(m_3+m_4)}+\mu_{34}\omega\frac{m_2(m_3+m_4)}{m_4(m_1+m_2)}]
\overrightarrow{Y}\cdot(\overrightarrow{X}-\overrightarrow{\xi})}.
\end{eqnarray}
\ruledown \vspace{0.5cm}
\begin{multicols}{2}
According to
\begin{eqnarray}
aY^2+bX^2+cX\cdot Y=a(Y+\frac{c}{2a}X)^2+(b-\frac{c^2}{4a})X^2,
\end{eqnarray}
we have
\end{multicols}
\ruleup
\begin{eqnarray}
\mathcal{V}_{13}(\xi)&=&(\frac{\mu_{12}\omega}{\pi})^{3/2}(\frac{\mu_{34}\omega}{\pi})^{3/2}\int
d\overrightarrow{X}\int d\overrightarrow{Y}\frac{e^{-mX}}{X}e^Me^N,
\end{eqnarray}
where
\begin{eqnarray}
M&=&-[\mu_{12}\omega(\frac{m_4}{m_3+m_4})^2+\mu_{34}\omega(\frac{m_2}{m_1+m_2})^2]
\left\{Y+\frac{\mu_{12}\omega\frac{m_4(m_1+m_2)}{m_2(m_3+m_4)}-\mu_{34}\omega
\frac{m_2(m_3+m_4)}{m_4(m_1+m_2)}}{2[\mu_{12}\omega(\frac{m_4}{m_3+m_4})^2+\mu_{34}\omega(\frac{m_2}{m_1+m_2})^2]}
(\overrightarrow{X}-\overrightarrow{\xi})\right\}^2\nonumber \\
\nonumber
&=&-[\mu_{12}\omega(\frac{m_4}{m_3+m_4})^2+\mu_{34}\omega(\frac{m_2}{m_1+m_2})^2]Y^{'2},\\
N&=&-\left\{\mu_{12}\omega(\frac{m_1+m_2}{2m_2})^2+\mu_{34}\omega(\frac{m_3+m_4}{2m_4})^2
-\frac{[\mu_{12}\omega\frac{m_4(m_1+m_2)}{m_2(m_3+m_4)}-\mu_{34}\omega
\frac{m_2(m_3+m_4)}{m_4(m_1+m_2)}]^2}{4[\mu_{12}\omega(\frac{m_4}{m_3+m_4})^2+\mu_{34}\omega(\frac{m_2}{m_1+m_2})^2]}\right\}
(\overrightarrow{X}-\overrightarrow{\xi})^2\nonumber \\
&=&-\frac{\mu_{12}\mu_{34}\omega}{\mu_{12}(\frac{m_4}{m_3+m_4})^2+\mu_{34}(\frac{m_2}{m_1+m_2})^2}(\overrightarrow{X}-\overrightarrow{\xi})^2.
\end{eqnarray}
\ruledown \vspace{0.5cm}
\begin{multicols}{2}
Furthermore, after making another coordinate transformation to Y as
\begin{eqnarray}
\nonumber
Y^{'}&=&\frac{\mu_{12}\omega\frac{m_4(m_1+m_2)}{m_2(m_3+m_4)}-\mu_{34}\omega
\frac{m_2(m_3+m_4)}{m_4(m_1+m_2)}}{2[\mu_{12}\omega(\frac{m_4}{m_3+m_4})^2+\mu_{34}\omega(\frac{m_2}{m_1+m_2})^2]}(\overrightarrow{X}-\overrightarrow{\xi})
\\&&+Y
\end{eqnarray}
and taking
\begin{eqnarray}
\alpha&=&\mu_{12}\omega(\frac{m_4}{m_3+m_4})^2+\mu_{34}\omega(\frac{m_2}{m_1+m_2})^2,\label{eq:07}
\\ \nonumber
\beta&=&\frac{\mu_{12}\mu_{34}\omega}{\mu_{12}(\frac{m_4}{m_3+m_4})^2+\mu_{34}(\frac{m_2}{m_1+m_2})^2},
\end{eqnarray}
we get
\end{multicols}
\ruleup
\begin{eqnarray}
\mathcal{V}_{13}(\xi)&=&(\frac{\mu_{12}\omega}{\pi})^{3/2}(\frac{\mu_{34}\omega}{\pi})^{3/2}\int
d\overrightarrow{X}\int
d\overrightarrow{Y}\frac{e^{-mX}}{X}e^{-\alpha
Y^2}e^{-\beta(\overrightarrow{X}-\overrightarrow{\xi})^2}
\end{eqnarray}
\ruledown\vspace{0.5cm}
\begin{multicols}{2}
Notice that
\end{multicols}
\ruleup
\begin{eqnarray}
\int d\overrightarrow{X}\frac{1}{X}
e^{-mX-\beta(\overrightarrow{X}-\overrightarrow{\xi})^2}
&=&e^{-\beta\xi^2}\int d\overrightarrow{X}\frac{1}{X}e^{-mX-\beta
X^2+2\overrightarrow{X}\cdot\overrightarrow{\xi}} \nonumber \\
&=&\frac{\pi\sqrt{\pi}}{2\beta\sqrt{\beta}\xi}e^{\frac{m^2}{4\beta}}\{e^{-m\xi}\{1-erf[-\sqrt{\beta}(\xi-\frac{m}{2\beta})]\}
-e^{m\xi}\{1-erf[\sqrt{\beta}(\xi+\frac{m}{2\beta})]\}\}\\
\int d\overrightarrow{Y}e^{-\alpha
Y^2}&=&\frac{\pi\sqrt{\pi}}{\alpha\sqrt{\alpha}}
\end{eqnarray}
\ruledown \vspace{0.5cm}
\begin{multicols}{2}
We finally get
\end{multicols}
\ruleup
\begin{eqnarray}
\mathcal{V}_{13}(\xi)&=&\frac{1}{2\xi}e^{\frac{m^2}{4\beta}}\{e^{-m\xi}\{1-erf[-\sqrt{\beta}(\xi-\frac{m}{2\beta})]\}
-e^{m\xi}\{1-erf[\sqrt{\beta}(\xi+\frac{m}{2\beta})]\}\},
\end{eqnarray}
and this is what we want in eq.(\ref{eq:08}).
 \ruledown \vspace{0.5cm}
\begin{multicols}{2}
\end{multicols}

\vspace{-1mm} \centerline{\rule{80mm}{0.1pt}}

\begin{multicols}{2}

\end{multicols}

 \clearpage

\end{document}